\documentclass[showpacs,preprintnumbers,10pt,twocolumn]{revtex4}%
\usepackage{amssymb}
\usepackage{amsfonts}
\usepackage{amsmath}
\usepackage{graphicx}
\usepackage{dcolumn}
\usepackage{bm}
\usepackage{revsymb}%
\ifx\pdfoutput\relax\let\pdfoutput=\undefined\fi
\newcount\msipdfoutput
\ifx\pdfoutput\undefined\else
\ifcase\pdfoutput\else
\ifx\paperwidth\undefined\else
\ifdim\paperheight=0pt\relax\else\pdfpageheight\paperheight\fi
\ifdim\paperwidth=0pt\relax\else\pdfpagewidth\paperwidth\fi
\fi\fi\fi
\begin{document}
\title{Universal resonancelike emergence of chaos in complex networks of
damped-driven nonlinear systems}
\author{$^{1}$Ricardo Chac\'{o}n and $^{2}$Pedro J. Mart\'{\i}nez}
\affiliation{$^{1}$Departamento de F\'{\i}sica Aplicada, E.I.I., Universidad de
Extremadura, Apartado Postal 382, E-06006 Badajoz, Spain and Instituto de
Computaci\'{o}n Cient\'{\i}fica Avanzada (ICCAEx), Universidad de Extremadura,
E-06006 Badajoz, Spain}
\affiliation{$^{2}$Departamento de F\'{\i}sica Aplicada, E.I.N.A., Universidad de Zaragoza,
E-50018 Zaragoza, Spain and Instituto de Nanociencia y Materiales de
Arag\'{o}n (INMA), CSIC-Universidad de Zaragoza, E-50009 Zaragoza, Spain}
\date{\today}

\begin{abstract}
Characterizing the emergence of chaotic dynamics of complex networks is an
essential task in nonlinear science with potential important applications in
many fields such as neural control engineering, microgrid technologies, and
ecological networks. Here, we solve a critical outstanding problem in this
multidisciplinary research field: The emergence and persistence of
spatio-temporal chaos in complex networks of damped-driven nonlinear
oscillators in the significant weak-coupling regime, while they exhibit
regular behavior when uncoupled. By developing a comprehensive theory with the
aid of standard analytical methods, a hierarchy of lower-dimensional effective
models, and extensive numerical simulations, we uncover and characterize the
basic physical mechanisms concerning both heterogeneity-induced and
impulse-induced emergence, enhancement, and suppression of chaos in starlike
and scale-free networks of periodically driven, dissipative nonlinear oscillators.

\end{abstract}

\maketitle

\textit{Introduction}.$-$Controlling the dynamical state of a complex network
is a fundamental problem in science [1-5] with many potential applications,
including neuronal [6] and ecological [7] networks. While most of these works
consider networks of linear systems [2,5], only lately has the generic and
richer case of networks of nonlinear systems [1,4] started to be investigated.
Also, the majority of studies of coupled nonlinear systems subjected to
external excitations focused on either local (homogeneous) diffusive-type or
global (all-to-all) coupling. However, little attention has been paid to the
possible influences of a heterogeneous connectivity on both the emergence and
strength of chaos in complex networks of \textit{nonautonomous} nonlinear
systems. Here we characterize the emergence and persistence (in parameter
space) of chaos in heterogeneous networks of damped-driven nonlinear systems
when the complex network presents a \textit{non-chaotic} state in the absence
of coupling, while a stable chaotic state emerges after coupling the same
nonautonomous nodes. Specifically, we study the \textit{interplay} among
heterogeneous connectivity, driving period, and impulse transmitted by a
homogeneous (non-harmonic) periodic excitation in the emergence and
persistence of spatio-temporal chaos in complex networks in the significant
weak-coupling regime. For the sake of clarity, the findings are discussed
through the analysis of starlike networks (SNs) of $N+1$ damped-driven
two-well Duffing oscillators. This system is sufficiently simple to obtain
analytical predictions while retaining the universal features of a dissipative
chaotic system. The complete model system reads%
\begin{align}
\overset{..}{x}_{H} &  =\left(  1-\lambda N\right)  x_{H}-x_{H}^{3}%
-\delta\overset{.}{x}_{H}+\gamma f\left(  t\right)  +\lambda\sum_{i=1}%
^{N}y_{i},\nonumber\\
\overset{..}{y}_{i} &  =\left(  1-\lambda\right)  y_{i}-y_{i}^{3}%
-\delta\overset{.}{y}_{i}+\gamma f\left(  t\right)  +\lambda x_{H},\tag{1}%
\end{align}
$i=1,...,N$, where $f(t)$ is a unit-amplitude $T$-periodic excitation and
$\lambda$ is the coupling. These equations describe the dynamics of a highly
connected node (or hub), $x_{H}$, and $N$ linked oscillators (or leaves),
$y_{i}$. For concreteness, we shall consider the elliptic excitation
$f(t)=f_{ellip}(t)\equiv$ $A(m)\operatorname{sn}(4Kt/T)\operatorname*{dn}%
\left(  4Kt/T\right)  $ [8] (see Fig. 1(a) and Supplemental Material (SM) [9]
for a detailed characterization of $f_{ellip}(t)$). In this Letter, we
concentrate on the relevant (typically asynchronous) case of sufficiently
small coupling, $\lambda$, external excitation amplitude, $\gamma$, and
damping coefficient, $\delta$, such that the dynamics of the leaves may be
decoupled from that of the hub on the one hand, and may be suitably described
as a periodic orbit around one of the potential minima on the other.
Specifically, we assume $\lambda=O\left(  \gamma^{2}\right)  $ throughout this work.

\begin{figure*}[ptb]
\centering
\includegraphics[width=0.23\textwidth]{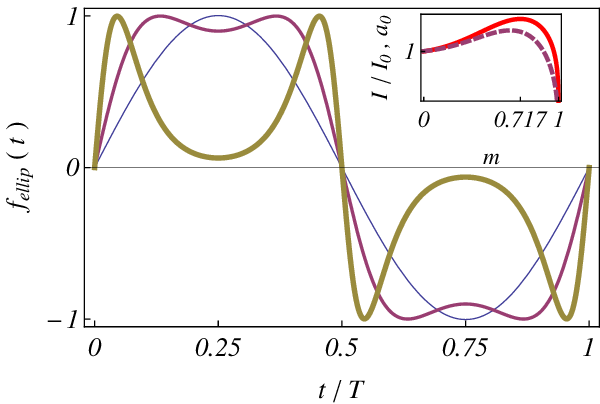}
\includegraphics[width=0.23\textwidth]{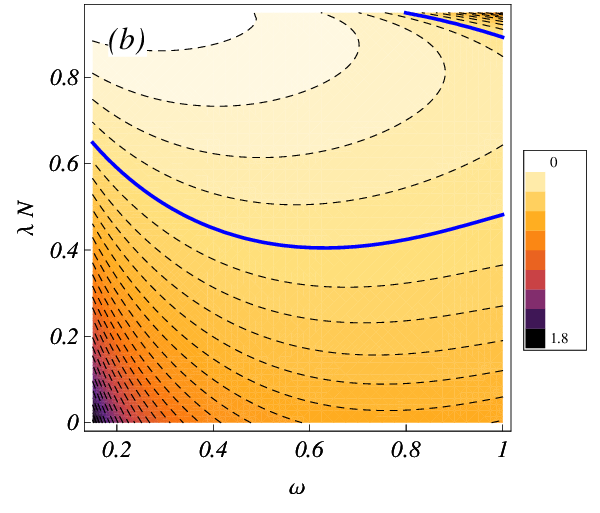}
\includegraphics[width=0.23\textwidth]{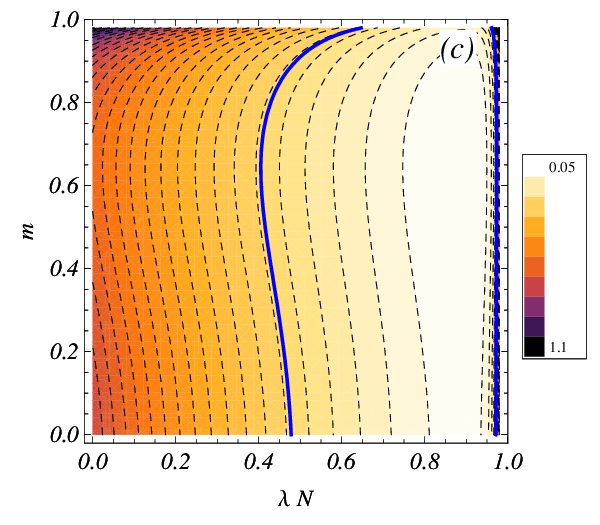}
\includegraphics[width=0.23\textwidth]{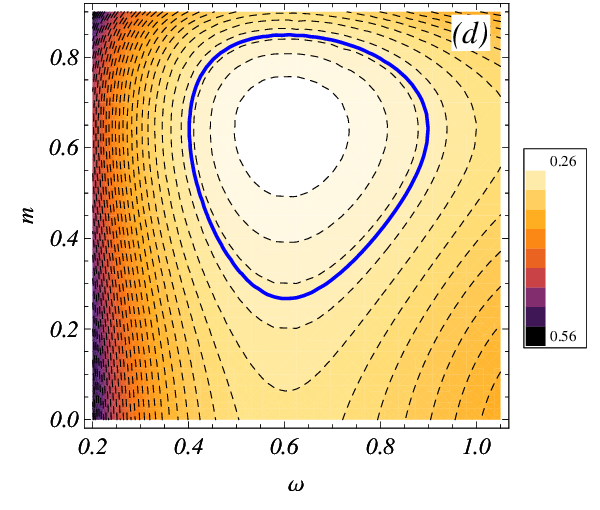} \caption{(a) External
excitation $f_{ellip}(t)=$ $A(m)\operatorname{sn}(4Kt/T)\operatorname*{dn}%
\left(  4Kt/T\right)  $ vs $t/T$, where $T$ is the period and $A(m)\equiv
1/\left\{  a+b/\left[  1+\exp\left(  \left\{  m-c\right\}  /d\right)  \right]
\right\}  $, with $a\equiv0.43932$, $b\equiv0.69796$, $c\equiv0.3727$, and
$d\equiv0.26883$, for three values of the shape parameter: $m=0$ (sinusoidal
pulse), $m=0.72\simeq m_{\max}$ (nearly square-wave pulse), and $m=0.999$
(double-humped pulse). Inset: The normalized impulse $I\left(  m\right)
/I\left(  m=0\right)  $ (solid line) and the Fourier coefficient $a_{0}(m)$
(dashed line; see the text). Chaotic threshold function $U\left(
\omega,\lambda N,m\right)  $ [cf. Eq.(7)] vs (b) $\omega$ and $\lambda N$ for
$m=0.65\simeq m_{\max}^{a_{0}}$, (c) $\lambda N$ and $m$ for $\omega=0.65$,
and (d) $\omega$ and $m$ for $\lambda N=0.45$. The solid blue lines indicate
the chaotic boundary corresponding to $\gamma/\delta=0.29$ [cf. Eq.(7)].}%
\label{fig1}%
\end{figure*}

\textit{ Effective model}.$-$Equations (1) for the leaves become%
\begin{equation}
\overset{..}{y}_{i}=y_{i}-y_{i}^{3}-\delta\overset{.}{y}_{i}+\gamma
f_{ellip}\left(  t\right)  , \tag{2}%
\end{equation}
$i=1,...,N$. After using the properties of the Fourier series of
$f_{ellip}\left(  t\right)  $ (see SM [9] for analytical details) and applying
standard perturbation methods [10] for the main resonance case, one obtains%
\begin{equation}
y_{i}\left(  t\rightarrow\infty\right)  \sim\xi_{i}+\frac{\gamma a_{0}\left(
m\right)  }{2-\omega^{2}}\sin\left(  \omega t\right)  , \tag{3}%
\end{equation}
where $\omega\equiv2\pi/T$, $\xi_{i}=\pm1$ depending on the initial
conditions, while the first Fourier coefficient of $f_{ellip}\left(  t\right)
$, $a_{0}\left(  m\right)  $, presents a single maximum at $m=m_{\max}^{a_{0}%
}\simeq0.65$ [see Fig 1(a), inset]. Since the initial conditions are randomly
chosen, this means that the quantities $\xi_{i}$ behave as discrete random
variables governed by Rademacher distributions. After inserting Eq. (3) into
Eq. (1), the resulting equation for the hub reads
\begin{equation}
\overset{..}{x}_{H}=\left(  1-\lambda N\right)  x_{H}-x_{H}^{3}-\delta
\overset{.}{x}_{H}+\Gamma\sin(\omega t)+\lambda\Xi, \tag{4}%
\end{equation}
where $\Xi\equiv\sum_{i=1}^{N}\xi_{i}$,$\ \Gamma\equiv\gamma a_{0}\left(
m\right)  \left[  1+\lambda N/\left(  2-\omega^{2}\right)  \right]  +O\left(
\gamma^{3}a_{0}^{2}\left(  m\right)  \right)  $. For finite $N$, the quantity
$\Xi$ behaves as a discrete random variable governed by a binomial
distribution with zero mean and variance $N$, while for sufficiently large $N$
one may assume that $\Xi$ behaves as a continuous random variable governed by
a normal distribution. Although the hub's dynamics are generally affected by
spatial quenched disorder through the term $\lambda\Xi$, one expects that it
may be neglected in the present case of weak coupling (WC) $\left(
1\gg\lambda\gtrsim0\right)  $ according to the above assumptions (see SM [9]
for a comparison of the cases with and without the term $\lambda\Xi$). Thus,
the network described by Eq. (1) can be \textit{effectively} replaced by a
hierarchy of reduced networks in which a hub is coupled to $M$ effective
leaves, each of which represents $n_{j}$ randomly chosen identical leaves
(i.e., leaves having exactly the same initial conditions) such that the
condition $\sum_{j=1}^{M}n_{j}=N$ is satisfied, in the WC regime and for
values of $m$ sufficiently less than $1$:%
\begin{align}
\overset{..}{x}_{H}  &  =\left(  1-\lambda N\right)  x_{H}-x_{H}^{3}%
-\delta\overset{.}{x}_{H}+\gamma a_{0}(m)\sin\left(  \omega t\right)
\nonumber\\
&  +\lambda\sum_{j=1}^{M}n_{j}y_{L,j},\nonumber\\
\overset{..}{y}_{L,j}  &  =\left(  1-\lambda\right)  y_{L,j}-y_{L,j}%
^{3}-\delta\overset{.}{y}_{L,j}+\gamma a_{0}(m)\sin\left(  \omega t\right)
+\lambda x_{H}, \tag{5}%
\end{align}
$j=1,...,M$, where $y_{L,j}$ represents the common leaf associated with each
group (cluster) of identical leaves. Equation (4) indicates that the
possibility of heterogeneity-induced emergence of chaos in the hub's dynamics
is now expected from the lowering of the potential barrier's height
$h\equiv\left(  1-\lambda N\right)  ^{2}/4$ as $N$ is increased on the one
hand, and the presence of the additional resonant excitation $\gamma
a_{0}\left(  m\right)  \lambda\left[  N/\left(  2-\omega^{2}\right)  \right]
\sin\left(  \omega t\right)  $ on the other. Notice that the amplitude of this
coupling-induced resonant excitation effectively depends upon the impulse
transmitted by $f_{ellip}\left(  t\right)  $ through the Fourier coefficient
$a_{0}\left(  m\right)  $. Quantitatively, this expectation can be deduced
with the aid of the Melnikov method (MM) [11,12]. Indeed, the application of
MM to Eq. (2) provides an estimate of a necessary condition for the emergence
of chaos:%

\begin{equation}
\frac{\gamma}{\delta}\geqslant U\left(  \omega,\lambda N=0,m\right)
\equiv\frac{2\sqrt{2}\cosh\left(  \pi\omega/2\right)  }{3\pi\omega a_{0}(m)},
\tag{6}%
\end{equation}
where $U\left(  \omega,\lambda N,m\right)  $ is the chaotic threshold
function. Assuming that $N$ satisfies the condition $0<\lambda N<1$ in order
to preserve the existence of an underlying separatrix for all $N$, the
application of MM to Eq. (4) after dropping the term $\lambda\Xi=O\left(
\gamma^{2}\right)  $ provides an estimate of the corresponding necessary
condition for the emergence of chaos:%
\begin{equation}
\frac{\gamma}{\delta}\geqslant U\left(  \omega,\lambda N,m\right)  \equiv
\frac{2\sqrt{2}\left(  1-\lambda N\right)  ^{3/2}\cosh\left(  \frac{\pi\omega
}{2\sqrt{1-\lambda N}}\right)  }{3\pi\omega a_{0}(m)\left(  1+\frac{\lambda
N}{2-\omega^{2}}\right)  } \tag{7}%
\end{equation}
(see SM [9] for a derivation of Eqs. (6) and (7)). Now, the following remarks
may be in order. First, the chaotic threshold function for the hub, Eq. (7),
reduces to that of the leaves, Eq. (6), when $\lambda N\rightarrow0$, i.e.,
for the limiting case of isolated nodes $\left(  \lambda=0\right)  $ and, in
the present WC regime, for the limiting case of homogeneous connectivity
$\left(  N=1\right)  $, as expected. Second, having fixed the ratio
$\gamma/\delta$ and the coupling $\lambda$, the possibility of chaotic
behaviour is predicted to be greater for the hub than for the leaves over wide
ranges of $\omega$, while this difference strongly depends on $N$ [cf. Eqs.
(6) and (7); see Fig. 1(b)]. Third, having fixed the coupling $\lambda$ and
the angular frequency $\omega$, the chaotic threshold function presents a
\textit{single minimum} in the $\lambda N-m$ parameter plane at $N=N_{\min
}\equiv N_{\min}(\omega)$ and $m=m_{\min}\simeq m_{\max}^{a_{0}}\simeq0.65$
(irrespective of the driving period), which means that the possibility of
chaotic behaviour is predicted to be higher when the impulse transmitted is
maximum and for intermediate values of $\lambda N$ than for the limiting cases
$\lambda N\rightarrow0,1$ [cf. Eq. (7); see Fig. 1(c)]. And fourth, having
fixed the coupling $\lambda$ and the number of leaves $N$, the chaotic
threshold function presents a \textit{single minimum} in the $\omega-m$
parameter plane at $\omega=\omega_{\min}\equiv\omega_{\min}(N)$ and
$m=m_{\min}\simeq m_{\max}^{a_{0}}\simeq0.65$ (irrespective of the driving
period), which means again that the possibility of chaotic behaviour is
predicted to be greater when the impulse transmitted is maximum and for
intermediate values of $\omega$ than for the limiting cases $\omega
\rightarrow0,\infty$ [cf. Eq. (7); see Fig. 1(d)]. Therefore, depending on the
remaining parameters, one could expect a heterogeneity-induced
(impulse-induced) route to chaos starting from a regular SN with a few leaves
(low-impulse excitation) by solely increasing their number (the excitation
impulse) on the one hand, and a heterogeneity-induced (impulse-induced) route
to regularity starting from a chaotic SN with many leaves (high-impulse
excitation) by further increasing (decreasing) their number (excitation
impulse) on the other.

Extensive numerical simulations of the complete system [Eq. (1)] and the
effective model system [Eq. (5)] confirmed an overall good agreement with
these expectations even for quite small values of $M$. Specifically, one can
compare the theoretical predictions and Lyapunov exponent (LE) calculations
[14] of both systems [Eqs. (1) and (5)] [15]. Illustrative examples are shown
in Figs. 2 and 3 for $N=138$ and values of $\delta,\gamma$ that are clearly
beyond the perturbative regime (compare Figs. 1(b), 1(c), 1(d) with Figs.
2(b), 2(a), 2(c), respectively). \begin{figure*}[ptb]
\centering
\includegraphics[width=0.3\textwidth]{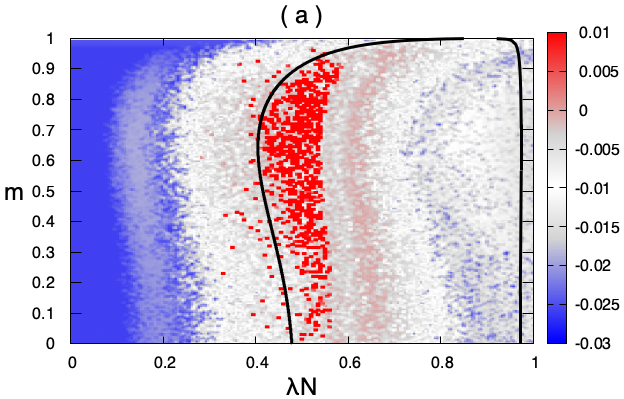}
\includegraphics[width=0.3\textwidth]{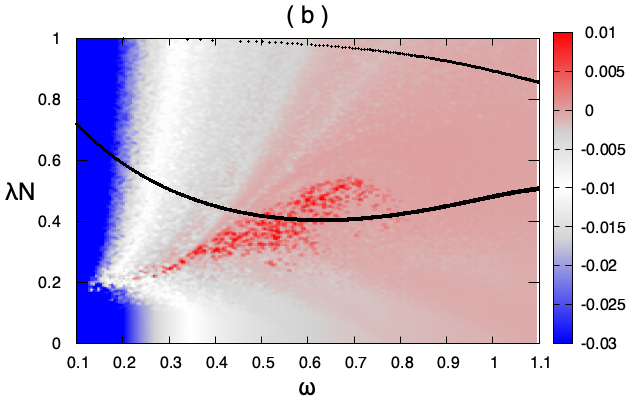}
\includegraphics[width=0.3\textwidth]{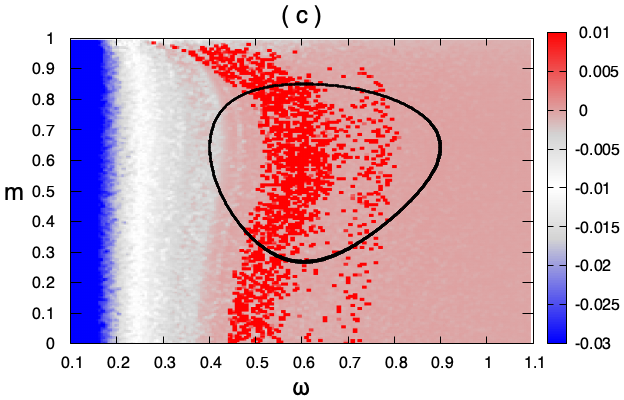}\\%
\includegraphics[width=0.3\textwidth]{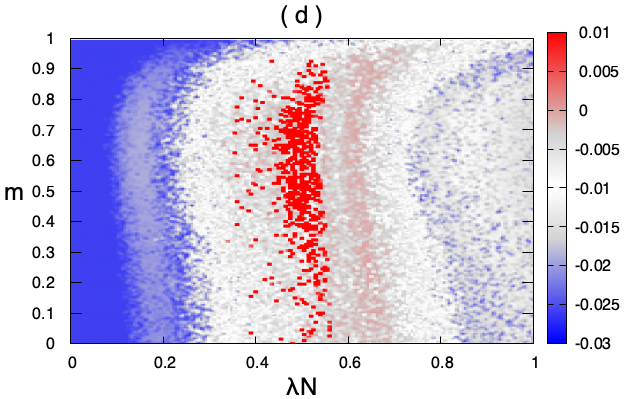}
\includegraphics[width=0.3\textwidth]{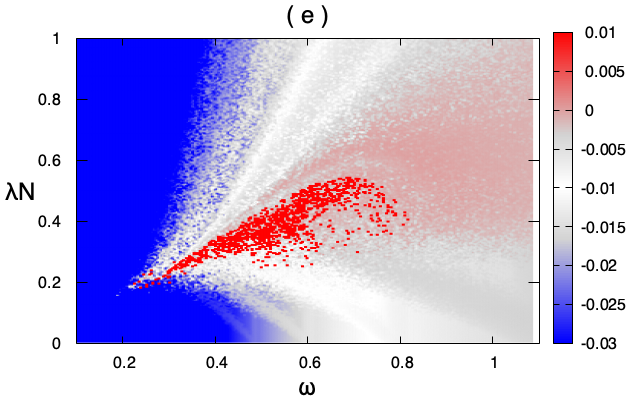}
\includegraphics[width=0.3\textwidth]{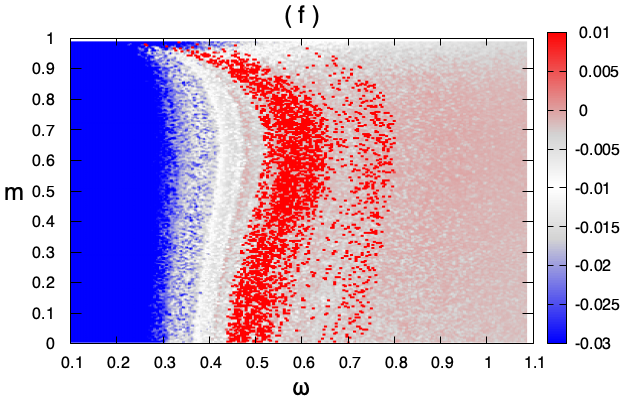} \caption{Maximal LE
distribution in the (a), (d) $\lambda N-m$, (b), (e) $\omega-\lambda N$, and
(c),(f) $\omega-m$ parameter planes for (a), (b), (c) the complete system [Eq.
(1)] and (d), (e), (f) the effective model system [Eq. (5) with $M=138$,
$n_{j}=1$, $j=1,...,138$] for (a), (d) $\omega=0.65$, (b), (e) $m=0.65\simeq
m_{\max}^{a_{0}}$, and (c), (f) $\lambda=0.00326$. Fixed parameters:
$N=138,\gamma=0.29,\delta=1$. The black lines indicate the chaotic boundary
corresponding to $\gamma/\delta=0.29$ [cf. Eq.(7)].}%
\label{fig2}%
\end{figure*}

\begin{figure*}[ptb]
\centering
\includegraphics[width=0.3\textwidth]{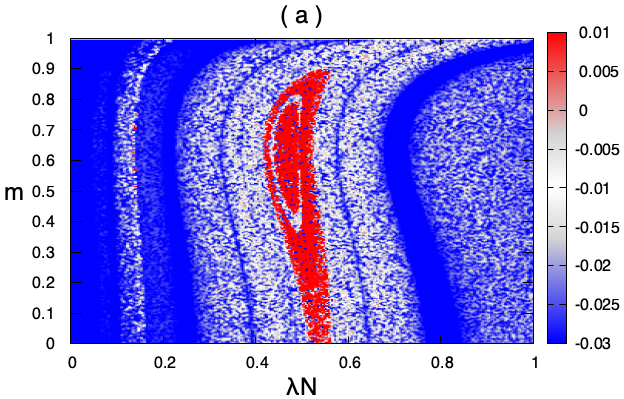}
\includegraphics[width=0.3\textwidth]{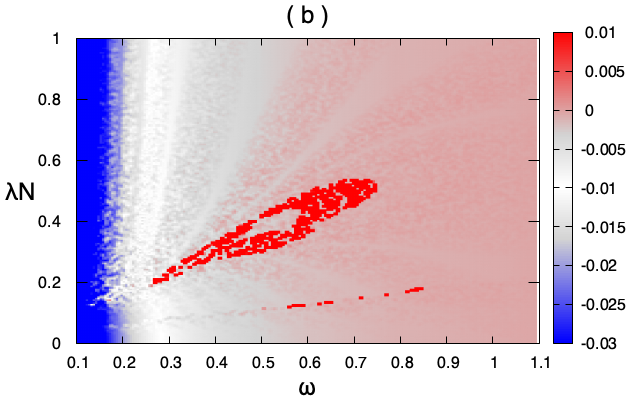}
\includegraphics[width=0.3\textwidth]{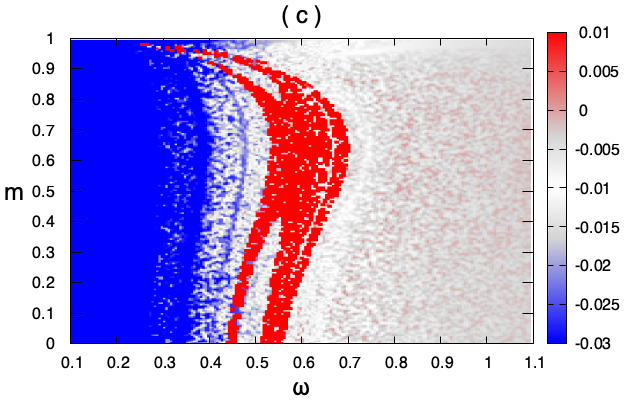}\\%
\includegraphics[width=0.3\textwidth]{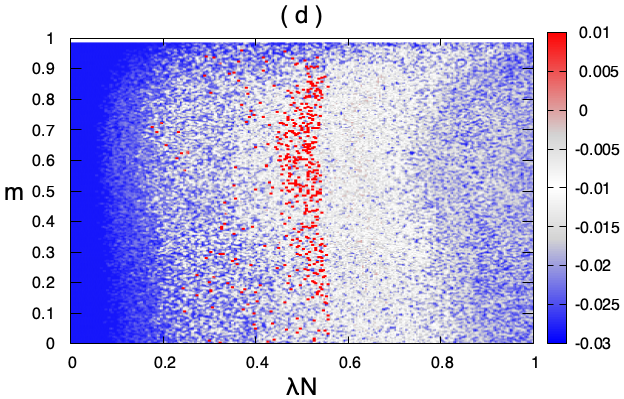}
\includegraphics[width=0.3\textwidth]{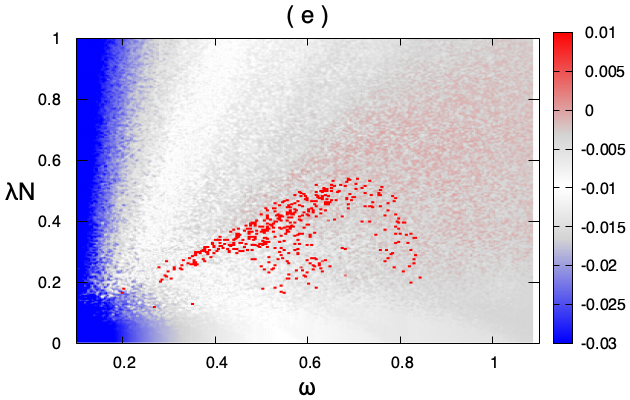}
\includegraphics[width=0.3\textwidth]{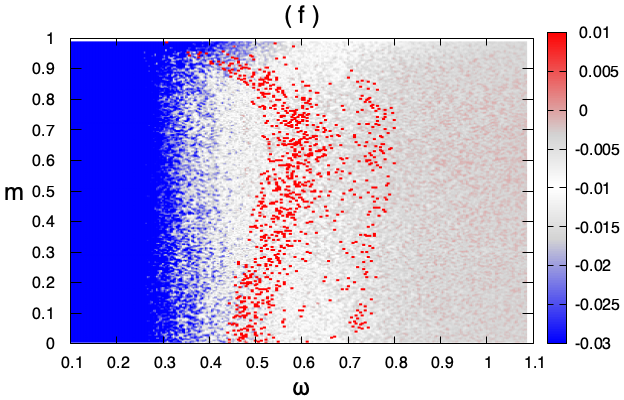} \caption{Maximal LE
distribution in the (a), (d) $\lambda N-m$, (b), (e) $\omega-\lambda N$, and
(c),(f) $\omega-m$ parameter planes for the effective model system [Eq. (5)]
for (a), (d) $\omega=0.65$, (b), (e) $m=0.65\simeq m_{\max}^{a_{0}}$, (c), (f)
$\lambda=0.00326$, and two values of the number of effective leaves: (a), (b),
(c) $M=2$ and (d), (e), (f) $M=24$. The values of $n_{j}$ were randomly chosen
while $\sum_{j=1}^{M}n_{j}=N$. Fixed parameters as in Fig. 2.}%
\label{fig3}%
\end{figure*}Typically, one finds for both systems [Eqs. (1) and (5)] a similar
resonancelike emergence of chaos in the $\lambda N-m$, $\omega-\lambda N$, and
$\omega-m$ parameter planes, which in its turn confirmed the effectiveness of
model Eq. (5), as is clearly seen when comparing Figs. 2(a), 2(b), 2(c) with
Figs. 2(d), 2(e), 2(f), respectively. As expected, the extent of the chaotic
regions is smaller in the case of the harmonic approximation $a_{0}%
(m)\sin\left(  \omega t\right)  $, which is due to the absence of the effects
of higher harmonics of $f_{ellip}\left(  t\right)  $. Remarkably, we found
that the emergence of chaos is attenuated and slightly distorted in the
parameter planes by decreasing the number $M$ of effective leaves from $M=N$,
as for the case $M=24$ shown in Figs. 3(d), 3(e), 3(f). This suppressory
effect occurs because the uniform initial randomness of the SN for $M=N$ is
broken as $M$ decreases from $N$ due to the formation of clusters of identical
leaves of \textit{different} cardinality, giving rise to an increase in
network desynchronization [16] which in turn makes it difficult to reach a
synchronized chaotic state. But, on restoring the uniformity of the initial
randomness, an increase in chaotic behaviour is observed even for quite small
values of $M$, such as for $M=2$ in which we took $n_{1}=n_{2}$ [cf. Figs.
3(a), 3(b), 3(c)].

\textit{Scale-free networks.}$-$Next, we discuss the possibility of extending
the results obtained for an SN to Barab\'{a}si-Albert (BA) networks [17] of
the same Duffing oscillators. The system is given by
\begin{equation}
\overset{..}{x}_{i}=x_{i}-x_{i}^{3}-\delta\overset{.}{x}_{i}+\gamma
f_{ellip}\left(  t\right)  -\lambda L_{ij}x_{j}, \tag{8}%
\end{equation}
$\ i=1,...,N$, where $L_{ij}=\kappa_{i}\delta_{ij}-A_{ij}$ is the Laplacian
matrix of the network, $\kappa_{i}=\sum_{j}A_{ij}$ is the degree of node $i$,
and $A_{ij}$ is the adjacency matrix with entries of $1$ if $i$ is connected
to $j$ and $0$ otherwise. Since in a BA network a highly connected node can be
thought of as a hub of a local SN with a certain degree $\kappa$ picked up
from the degree distribution ($P(\kappa)\sim\kappa^{-\alpha}$), one could
expect the above scenario for SNs to remain valid to some degree. Indeed, for
each hub with a sufficiently high (depending on the remaining parameters)
degree $\kappa_{i}$, one systematically observes that the bifurcation diagram
of its velocity $\overset{.}{x}_{i}$ vs coupling $\lambda$ presents,
essentially, the same overall chaotic window over the range $0<\lambda
\kappa_{i}<1$, in accordance with the predictions from the above SN scenario
(see Figs. 4(b)). This is reflected in both the global chaos of the BA
network, as shown in Fig. 4(a), and the number of chaotic nodes of the
network, $N_{chaos}$, as shown in Fig. 4(c). When $\lambda\kappa_{i}%
\geqslant1$, the potential associated with each hub of degree $\kappa_{i}$
undergoes a topological change, thus preventing the emergence of homoclinic
chaos in such a hub. Therefore, for $\lambda$ values sufficiently far from the
WC regime, the emergence of chaos in the BA network is no longer possible, as
is confirmed by LE calculations (see Fig. 4(c) and SM [9] for additional examples).

\begin{figure*}[ptb]
\centering
\includegraphics[width=0.3\textwidth]{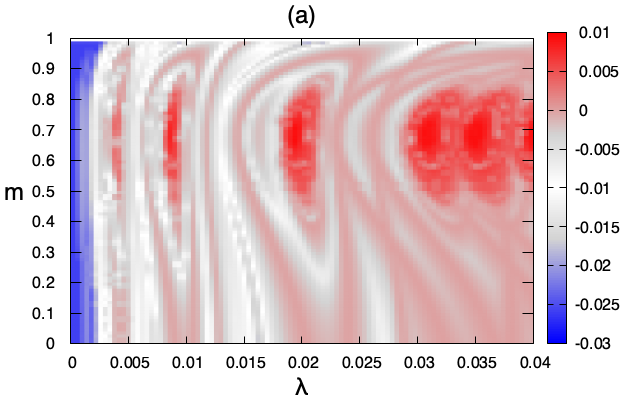}
\includegraphics[width=0.3\textwidth]{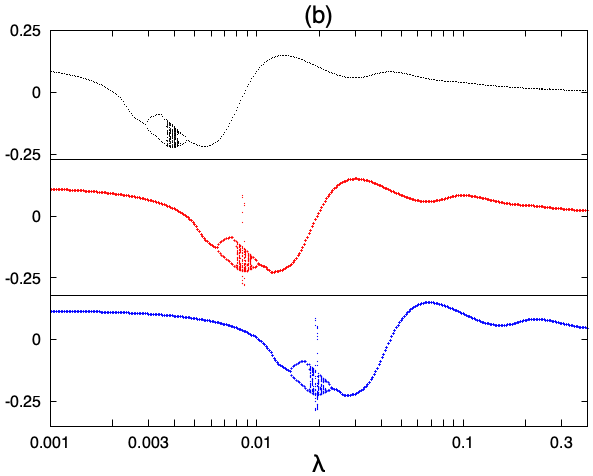}
\includegraphics[width=0.3\textwidth]{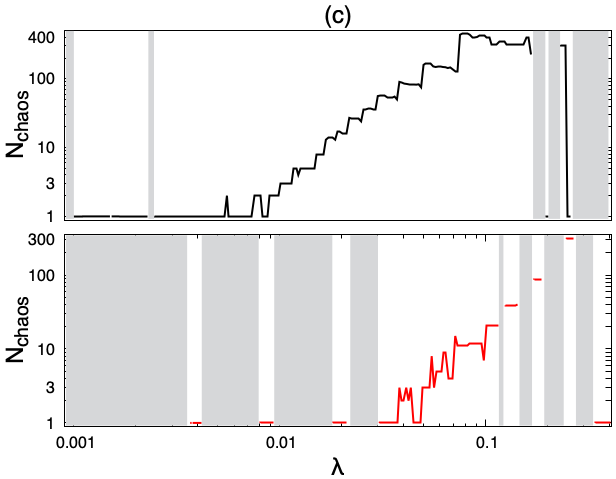} \caption{ (a) Maximal LE
distribution in the $\lambda-m$ parameter plane for a scale-free network [Eq.
(8)] with $N=500,\alpha=2.7$, $\gamma=0.29,\delta=1,T=2\pi/0.65$. (b)
Bifurcation diagrams for the velocities $\overset{.}{x}_{i}$ of its three main
hubs ($\kappa_{1}=138$, $\kappa_{2}=61$, $\kappa_{3}=27$, from top to bottom)
vs coupling $\lambda$ (logarithmic scale) for $m=0.65\simeq m_{\max}^{a_{0}}$.
(c) Number of chaotic nodes $N_{chaos}$ (logarithmic scale) vs coupling
$\lambda$ (logarithmic scale) for two sets of parameters $\left(
\gamma,\delta,T\right)  $: $(0.2,0.154,2\pi/0.5)$ (top) and ($0.29,1,2\pi
/0.65)$ (bottom), and the remaining parameters as in (a) (grey regions denote
$N_{chaos}=0$).}%
\label{fig4}%
\end{figure*}

\textit{Conclusion.}$-$Basic physical mechanisms have been discussed
concerning both heterogeneity-induced and impulse-induced emergence,
enhancement, and suppression of chaos in complex networks of periodically
driven, dissipative nonlinear systems in the significant weak-coupling regime.
With the aid of a hierarchy of lower-dimensional effective models and
extensive numerical simulations, we have characterized the resonancelike
interplay among heterogeneous connectivity, impulse transmitted by a
homogeneous periodic excitation, and its driving period in the emergence and
persistence of spatio-temporal chaos in starlike and scale-free networks of
bistable oscillators. In view of the simplicity and generality of this
multiple resonancelike scenario and the great robustness and scope of the
physical mechanisms involved, we expect it to be quite readily testable by
experiment, for instance in the context of nonlinear electronic circuits.
Finally, we hope our results can serve as an important step towards
understanding emergence of chaos in complex networks of interconnected
damped-driven nonlinear systems in the case of time-varying connections [18],
while the exploration of both the effectiveness of local application of
additional chaos-suppressing excitations and of the effects of different
coupling functions [19] represent exciting next steps for future research.

\begin{acknowledgments}
R.C. acknowledges financial support from the Ministerio de Ciencia e
Innovaci\'{o}n (MICINN, Spain) through Project No.
PID2019-108508GB-I00/AEI/10.13039/501100011033 cofinanced by FEDER funds.
P.J.M. acknowledges financial support from the Ministerio de Ciencia e
Innovaci\'{o}n (MICINN, Spain) through Project No.
PID2020-113582GB-I00/AEI/10.13039/501100011033 cofinanced by FEDER funds and
from the Gobierno de Arag\'{o}n (DGA, Spain) through Grant No. E36\_23R.
\end{acknowledgments}


\begin{thebibliography}{99}
\bibitem {13}Regarding numerical experiments, this means the appearance of
chaotic transients but not necessarily of steady chaos (see, e.g., Ref. [12]).
\bibitem {15}Notice the caveat that one cannot expect too good a quantitative
agreement between the two kinds of results because LE provides information
concerning steady chaos, while MM is a perturbative method [13].
\end{thebibliography}
\end{document}


\title{Supplemental Material\\Universal resonancelike emergence of chaos in starlike networks of
damped-driven nonlinear systems}
\author{Ricardo Chac\'{o}n$^{1}$ and Pedro J. Mart\'{\i}nez$^{2}$}
\affiliation{$^{1}$Departamento de F\'{\i}sica Aplicada, E.I.I., Universidad de
Extremadura, Apartado Postal 382, E-06006 Badajoz, Spain, and Instituto de
Computaci\'{o}n Cient\'{\i}fica Avanzada (ICCAEx), Universidad de Extremadura,
E-06006 Badajoz, Spain}
\affiliation{$^{2}$Departamento de F\'{\i}sica Aplicada, E.I.N.A., Universidad de Zaragoza,
E-50018 Zaragoza, Spain and Instituto de Nanociencia y Materiales de
Arag\'{o}n (INMA), CSIC-Universidad de Zaragoza, E-50009 Zaragoza,
Spain}
\maketitle

\begin{center}

\end{center}

{\LARGE \bigskip}

\section{Theoretical methods}

\subsection{Fourier expansion of the periodic elliptic excitation}

In our study we consider the elliptic excitation
\begin{equation}
f(t)\equiv A\operatorname*{sn}\left(  4Kt/T\right)  \operatorname*{dn}\left(
4Kt/T\right)  ,\tag{S1}%
\end{equation}
in which $\operatorname*{sn}\left(  \cdot\right)  \equiv\operatorname*{sn}%
\left(  \cdot;m\right)  $ and $\operatorname*{dn}\left(  \cdot\right)
\equiv\operatorname*{dn}\left(  \cdot;m\right)  $ are Jacobian elliptic
functions of parameter $m$ ($K\equiv K(m)$ is the complete elliptic integral
of the first kind) [1] and
\begin{equation}
A=A(m)\equiv\left[  a+b\left(  1+\exp\left\{  \frac{m-c}{d}\right\}  \right)
^{-1}\right]  ^{-1},\tag{S2}%
\end{equation}
is a normalization function ($a\equiv0.43932,b\equiv0.69796,c\equiv
0.3727,d\equiv0.26883$) which is introduced for the elliptic excitation to
have the same amplitude, 1, and period $T$, for any waveform (i.e., $\forall
m\in\left[  0,1\right]  $). When $m=0$, then $f\left(  t\right)  _{m=0}%
=\sin\left(  2\pi t/T\right)  $, i.e., one recovers the standard case of an
harmonic excitation, while for the limiting value $m=1$ the excitation
vanishes. The effect of renormalization of the elliptic arguments is clear:
with $T$ constant, solely the excitation's impulse is varied by increasing the
shape parameter $m$ from $0$ to $1$. Note that, as a function of $m$, the
elliptic excitation's impulse per unit of period
\begin{equation}
I=I(m)\equiv\frac{N\left(  m\right)  }{2K\left(  m\right)  }\tag{S3}%
\end{equation}
presents a single maximum at $m=m_{\max}^{impulse}\simeq0.717$ (see Fig. S1).

The Fourier expansion of the elliptic excitation (Eq. S1) reads%
\begin{align}
f(t) &  =\sum_{n=0}^{\infty}a_{n}(m)\sin\left[  \left(  2n+1\right)  \left(
\frac{2\pi t}{T}\right)  \right]  ,\tag{S4}\\
a_{n}(m) &  \equiv\frac{\pi^{2}A(m)(n+\frac{1}{2})}{\sqrt{m}K^{2}%
(m)}\operatorname{sech}\left[  \frac{(n+\frac{1}{2})\pi K(1-m)}{K(m)}\right]
,\tag{S5}%
\end{align}
in which its Fourier coefficients satisfy the properties: i) $\lim
_{m\rightarrow1}a_{n}(m)=0$, ii) $a_{n}(m)$ exhibits a single maximum at
$m=m_{\max}\left(  n\right)  $ such that $m_{\max}\left(  n+1\right)
>m_{\max}\left(  n\right)  $, $n=0,1,...$, iii) the normalized functions
$a_{0}(m)/a_{0}(m=0)$ and $I(m,T)/I(m=0,T)\equiv\pi$ $A(m)/(2K(m))$ present,
as functions of $m$, similar behaviours while their maxima verify that
$m_{\max}\left(  n=0\right)  \simeq0.642$ is very close to $m_{\max}%
^{impulse}\simeq0.717$ (see Fig. S1), and iv) the Fourier expansion (Eq. S4)
is rapidly convergent over a wide range of values of the shape parameter. The
following remarks may now be in order. First, regarding analytical estimates,
the properties (iii) and (iv) are relevant in the sense that they allow us to
obtain an useful effective estimate of the chaotic threshold in parameter
space from Melnikov method (MM) [2,3] by solely retaining the first harmonic
of the Fourier expansion (Eq. S4):%
\begin{equation}
f(t)\approx a_{0}(m)\sin\left(  \omega t\right)  ,\tag{S6}%
\end{equation}
$\omega\equiv2\pi/T$. Second, regarding numerical simulations, we considered
the entire Fourier expansion of the elliptic excitation in order to obtain
useful information concerning the effectiveness of the approximations used in
the theoretical analysis.

\begin{figure}[ptb]
\includegraphics[width=0.4\textwidth]{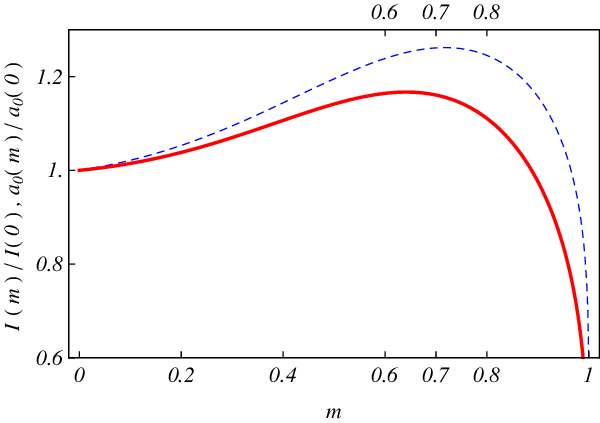}\caption{Normalized first
Fourier coefficient $a_{0}(m)/a_{0}(m=0)$ (Eq. S5, solid line) and elliptic
excitation's impulse $I(m)/I(m=0)\equiv\pi$ $A(m)/(2K(m))$ (Eq. S3, dashed
line) versus shape parameter $m$. We can see that the respective single maxima
occur at very close values of the shape parameter: $m_{\max}\left(
n=0\right)  \simeq0.642$ and $m_{\max}^{impulse}\simeq0.717$, respectively. }%
\label{figs1}%
\end{figure}

\subsection{Chaotic threshold from Melnikov method}

Melnikov introduced a function (now known as the Melnikov function (MF),
$M\left(  t_{0}\right)  $) which measures the distance between the perturbed
stable and unstable manifolds in the Poincar\'{e} section at $t_{0}$. Although
the predictions from MM are both limited (only valid for motions based at
points sufficiently near the separatrix) and approximate (the MM is a
first-order perturbative method), they are of great interest due to the
general scarcity of such analytical results in the theory of chaos. Since it
has been described many times by distinct authors [3], we do not discuss it in
detail here, but analyze the results obtained from it. Regarding Eqs. (2) and
(4) in the main text, note that keeping with the assumption of the MM, it is
assumed that the amplitudes of the dissipation and excitation terms are
sufficiently small $\left(  0<\delta,\gamma\ll1\right)  $. The application of
MM to Eq. (2) in the main text give us the Melnikov function (MF)%
\begin{equation}
M_{0L}^{\pm}(t_{0})=-C\pm A\sin\left(  \omega t_{0}\right)  ,\tag{S7}%
\end{equation}
with%
\begin{align}
C &  \equiv4\delta/3,\nonumber\\
A &  \equiv\sqrt{2}\pi\gamma a_{0}(m)\omega\operatorname{sech}\left(
\pi\omega/2\right)  ,\tag{S8}%
\end{align}
where the positive (negative) sign refers to the right (left) homoclinic orbit
(of the underlying conservative system):
\begin{align}
y_{i,0}(t) &  =\pm\sqrt{2}\operatorname{sech}\left(  t\right)  ,\tag{S9}\\
\overset{.}{y}_{i,0}\left(  t\right)   &  =\mp\sqrt{2}\operatorname{sech}%
\left(  t\right)  \tanh\left(  t\right)  .\nonumber
\end{align}
If the MF $M_{0L}^{\pm}(t_{0})$ has a simple zero, then a homoclinic
bifurcation occurs, signifying the \textit{onset} of chaotic instabilities
[3]. Clearly, the MF (S7) has simple zeros when%
\begin{equation}
\frac{\gamma}{\delta}\geqslant U\left(  \omega,\lambda N=0,m\right)
\equiv\frac{2\sqrt{2}}{3\pi\omega a_{0}(m)}\cosh\left(  \pi\omega/2\right)
,\tag{S10}%
\end{equation}
where the equals sign corresponds to the case of tangency between the stable
and unstable manifolds, while $U\left(  \omega,\lambda N,m\right)  $ is the
chaotic threshold function (Eq. (6) in the main text). Assuming that $N$
satisfies the condition $0<\lambda N<1$ in order to preserve the existence of
an underlying separatrix for all $N$, the application of MM to Eq. (4) in the
main text after dropping the term $\lambda\Xi=O\left(  \gamma^{2}\right)  $
yields the MF
\begin{equation}
M_{0H}^{\pm}(t_{0})=-C^{\prime}\pm A^{\prime}\cos\left(  \omega t_{0}\right)
,\tag{S11}%
\end{equation}
with
\begin{align}
C^{\prime} &  \equiv4\delta\left(  1-\lambda N\right)  ^{3/2}/3,\nonumber\\
A^{\prime} &  \equiv\sqrt{2}\pi\Gamma\omega\operatorname{sech}\left[
\pi\omega/\left(  2\sqrt{1-\lambda N}\right)  \right]  .\tag{S12}%
\end{align}
The MF $M_{0H}^{\pm}(t_{0})$ has simple zeros when%
\begin{equation}
\frac{\gamma}{\delta}\geqslant U\left(  \omega,\lambda N,m\right)  \equiv
\frac{2\sqrt{2}\left(  1-\lambda N\right)  ^{3/2}\cosh\left(  \frac{\pi\omega
}{2\sqrt{1-\lambda N}}\right)  }{3\pi\omega a_{0}(m)\left(  1+\frac{\lambda
N}{2-\omega^{2}}\right)  }.\tag{S13}%
\end{equation}
which is Eq. (7) in the main text. Thus, the chaotic boundary in parameter
space is given by%
\begin{equation}
\frac{\gamma}{\delta}=U\left(  \omega,\lambda N,m\right)  .\tag{S14}%
\end{equation}

\section{Aditional numerical results}

Figure S2 shows the effect of the spatial quenched disorder through the term
$\lambda\Xi$ [cf. Eq. (4) in the main text]. One sees an overall agreement
between the cases with and without the term $\lambda\Xi$.

\begin{figure}[htb]
\centering
\includegraphics[width=0.35\textwidth]{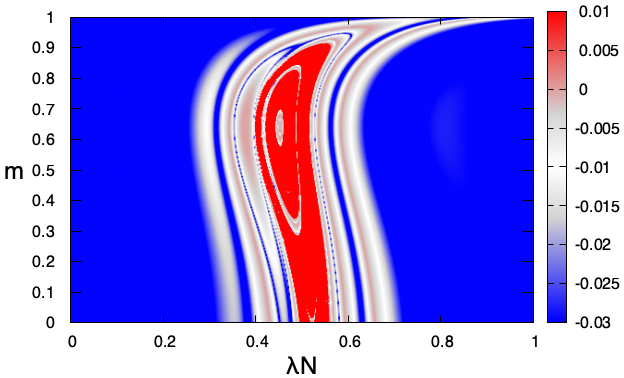}
\includegraphics[width=0.35\textwidth]{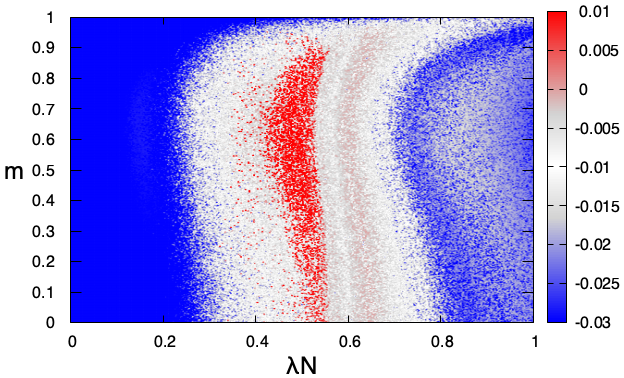}
\includegraphics[width=0.25\textwidth]{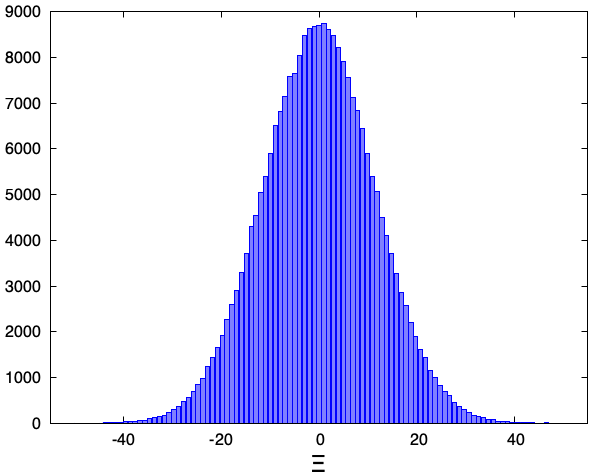}
\caption{Maximal
Lyapunov exponent distribution in the $\lambda N-m$ parameter plane for the
case (a) with and (b) without the term $\lambda\Xi$ in Eq. (4) in the main
text and $N=138,\delta=1,\gamma=0.29,T=2\pi/0.65$. (c) Histogram of the
quantity $\Xi$ indicating a normal distribution. The data shown is a random
sample of $20,{}100$ points.}%
\label{figS2}%
\end{figure}

\begin{figure}[htb]
\centering
\includegraphics[width=0.6\textwidth]{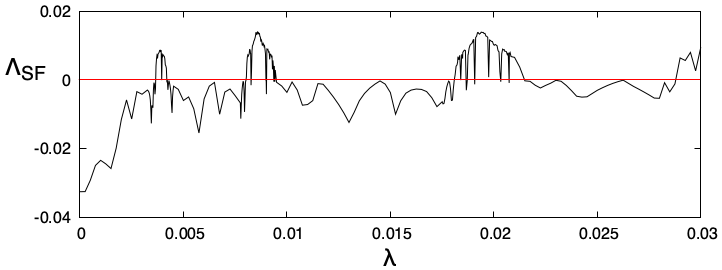}\\
\includegraphics[width=0.6\textwidth]{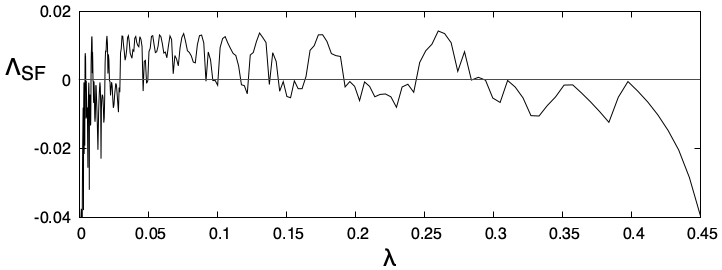}
\caption{Maximal
Lyapunov exponent of a scale-free network, $\Lambda_{SF}$, versus the coupling
$\lambda$ over two ranges: $\lambda\in\left[  0,0.03\right]  $ (top) and
$\lambda\in\left[  0,0.45\right]  $ (bottom). Fixed parameters: $N=500,\alpha
=2.7$, $\gamma=0.29,\delta=1,T=2\pi/0.65$.}%
\label{figS3}%
\end{figure}

Figure S3 shows the maximal Lyapunov exponent of a scale-free network,
$\Lambda_{SF}$, versus the coupling $\lambda$ for the same parameters as in
Fig. 4(a) in the main text. One sees that the appearance of chaos in the
network is exclusively due to the emergence of chaos in the most connected
hubs in the weak coupling regime (see Fig. S3(a)), while the broadening of the
chaotic windows beyond the weak coupling regime is due to the activation of
chaos in a multitude of lower degree nodes (see Fig. S3(b)), as expected from
the power-law distribution $P(\kappa)\sim\kappa^{-\alpha}$.
